\documentclass[pra,floatfix,twocolumn,superscriptaddress]{revtex4-1}
\usepackage[final]{graphicx}
\usepackage{times,bbm,amsmath,amssymb}
\usepackage{epsfig,color}
\usepackage{hyperref}
\usepackage{float,siunitx}
\usepackage[caption = false]{subfig}
\usepackage[greek,english]{babel}
\usepackage{thumbpdf,enumerate}
\usepackage{booktabs}
\usepackage{sidecap}
\usepackage[scaled=.8]{couriers}
\usepackage{pstricks}
\usepackage{multirow}
\usepackage{placeins}
\usepackage{pst-grad}
\usepackage{epigraph}
\usepackage{longtable}
\usepackage{booktabs}
\usepackage{gensymb}

\begin{document}

\title{Spectral-gap immune characterization of electric fields}

\author{Ilaria Gianani}\email{ilaria.gianani@gmail.com}
\affiliation{Clarendon Laboratory, University of Oxford, Parks Road, Oxford OX1 3PU, UK}

\author{Charles Bourassin-Bouchet}
\affiliation{Clarendon Laboratory, University of Oxford, Parks Road, Oxford OX1 3PU, UK}
\affiliation{Synchrotron SOLEIL, Université Paris-Saclay, Saint-Aubin, BP 34, 91192 Gif-sur-Yvette, France}

\author{Patrick N. Anderson}
\affiliation{Clarendon Laboratory, University of Oxford, Parks Road, Oxford OX1 3PU, UK}

\author{Matthias M. Mang}
\affiliation{Clarendon Laboratory, University of Oxford, Parks Road, Oxford OX1 3PU, UK}

\author{Adam S. Wyatt}
\affiliation{Central Laser Facility, STFC Rutherford Appleton Laboratory, Harwell OX11 0QX, UK}

\author{Marco Barbieri} 
\affiliation{Dipartimento di Scienze, Universit\`a degli Studi Roma Tre, Via della Vasca Navale 84, 00146, Rome, Italy}

\author{Ian A. Walmsley}
\affiliation{Clarendon Laboratory, University of Oxford, Parks Road, Oxford OX1 3PU, UK}

\begin{abstract}
Synthesised light sources need reliable diagnostics for effective application to sub-femtosecond control and probing. However, commonly employed self-referencing techniques for pulsed-field characterisation fail in the presence of wide spectral gaps, while direct sampling methods are limited to high intensities. Here, we introduce a new approach labelled SPectral-gap Immune Characterisation of Electric Fields (SPICE), which overcomes these barriers by means of
multi-spectral shearing interferometry, using an unknown reference and a reconstruction algorithm that makes use of this redundant information to simultaneously reconstruct the reference and test pulse with high precision and accuracy. We envisage that this technique will help foster new applications for  broadband sources.
\end{abstract}

\maketitle

\section{Introduction.}
The study and control of dynamical phenomena occurring at the molecular and atomic time scale prompted, over the last few decades, the developement of ultrafast light sources \cite{Walmsley2009,Krausz2009}.
The ability to shape an optical pulse in order to perform a specific task, such as excitation of a particular species or process, or to produce a particularly brief pulse for probing the most rapid dynamical phenomena, demands as a corollary the ability to characterise the pulse. Nowadays, self-referencing techniques offer viable characterization methods in the femtosecond range, and this has made applications such as the control of chemical reaction possible~\cite{Potter,Brumer1986}.
The introduction of light-wave synthesizers open up the possibility of scaling these capabilities to the sub-femtosecond regime \cite{Manzoni2015}, provided we deploy adequate pulse shaping and characterization techniques. 
Light-wave sources include schemes based on pulse broadening in hollowcore fibre \cite{Wirth2011}, frequency combs, or optical parametric oscillators \cite{McCracken2012, Huang2011}. 
In many instances, these sources show a spectral structure composed of well-separated spectral components. Standard self-referenced or cross-referenced techniques are unable to reconstruct these waveforms. Multi-shearing is an effective technique for characterizing the spectral phase in the presence of gaps in the spectrum narrower than the width of the peaks \cite{Austin2010b}. This cannot be directly extended to the regime where the gaps are broader than the peaks as the device will fail to reconstruct the relative phases between the spectral peaks. A possible solution might be to adopt a referenced scheme such as X-SPIDER \cite{Hirasawa2002,Morita2002} or X-FROG \cite{Linden1998}, but in these geometries the challenge is to obtain a well characterized reference, a task which can be difficult, especially for extremely large bandwidths, and the results may also be prone to ambiguities \cite{Keusters2003,Seifert2004}. Recent progress has been made in this area using a new spectrographic technique (VAMPIRE) \cite{Seifert2009} which alleviates the blind-FROG ambiguities \cite{Seifert2004} by introducing a conditioning filter and applying an iterative algorithm. As an alternative strategy High-Harmonic Generation (HHG) can be used for the direct sampling of the electric field \cite{ARIES,Kim2013} , however this imposes the requirement of high intensities and a complex apparatus which is often unavailable nor necessary in many applications.

In this paper we experimentally demonstrate that a direct, unambiguous reconstruction of the spectral phase can be achieved by means of a new method of spectral shearing interferometry. We overcome the barriers associated with the characterization of an external reference by adopting a scheme based on SEA-CAR SPIDER \cite{Austin2010b,Witting2009a} combined with the recent Mutual Interferometric Characterization of independent Electric fields (MICE) algorithm \cite{BourassinBouchet2013} that utilizes redundant information present in the multi-shear scheme to simultaneously characterize both the reference and test pulse. We call our technique SPectral-gap Immune Characterization of Electric fields (SPICE). 
 A useful feature of our spectral interferometric approach is that it can be applied to lower energy pulses than HHG-based sampling \cite{ARIES} or the self-referencing spectrographic equivalent VAMPIRE \cite{Seifert2009}s, since it uses an independent, uncharacterized pulse as a reference.



\section{The SPICE method}

In order to illustrate the principle of our technique we consider the simplest case of a test pulse (TP) with a spectrum consisting of two peaks, separated by a gap larger than the spectral width of each of the individual peaks. In order to successfully reconstruct the spectral phase of such a TP, we must reconstruct both the phase within each peak and the relative phase between the peaks. Conventional techniques enable the former to be estimated, but measuring the latter remains a largely unsolved problem. In order to retrieve the relative phase, an innovative way to infer phase information despite the presence of spectral nulls must be introduced. This is possible by preparing the ancillary pulses involved in the characterization to "bridge" the gap. This forms the core of our three-step SPICE method.

\begin{figure}[t]
\centering
\includegraphics[width=0.45\textwidth]{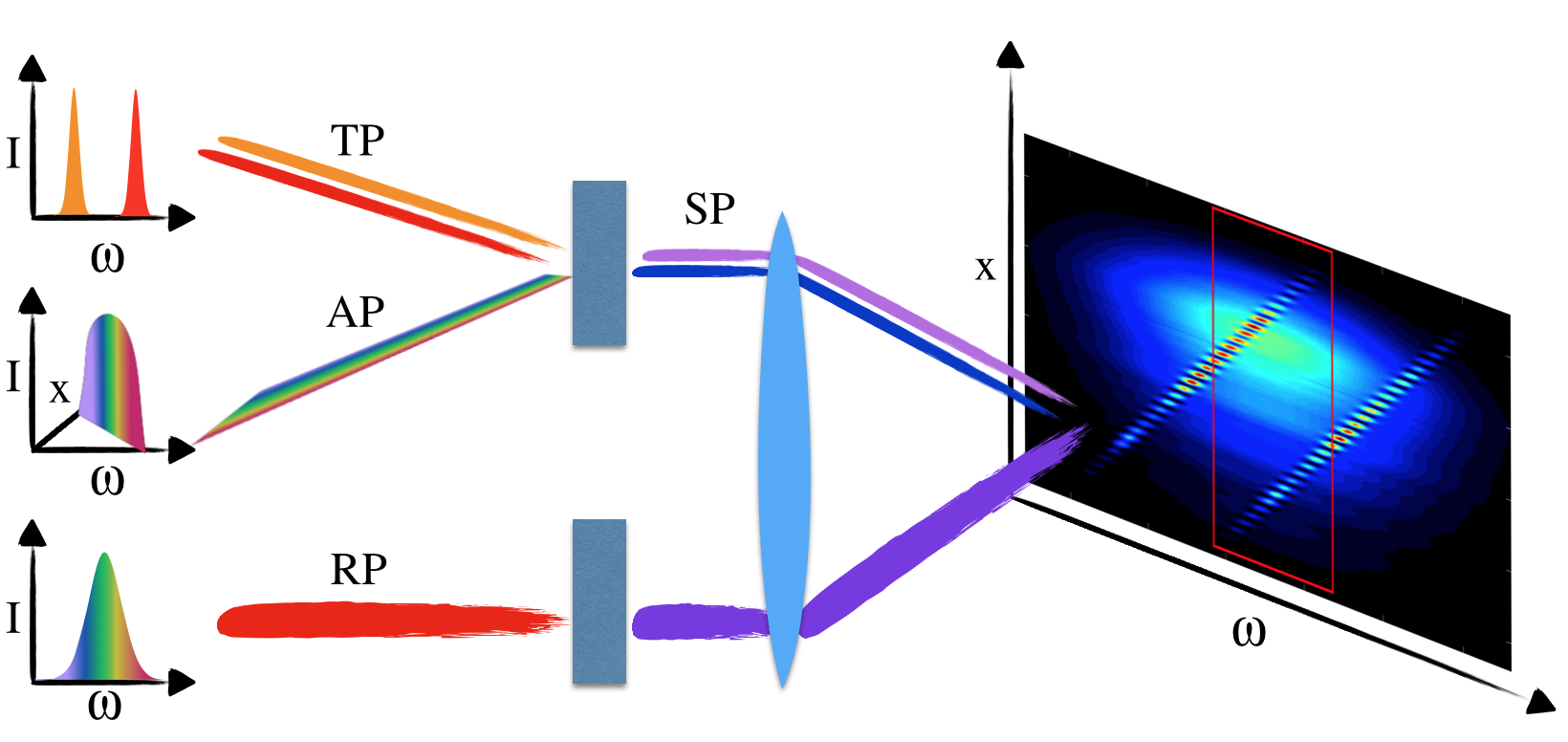}
\caption{ The SPICE concept. First, a test pulse (TP) with a large spectral gap is upconverted with a spatially-chirped ancillary pulse (AP). Next, this is interfered with an independent reference pulse (RP) at the entrance slit to an imaging spectrometer. The resulting spatio-spectral (2D) interferogram is processed by the MICE algorithm, yielding the phase information of the three pulses. See the text for a detailed discussion of the bandwidth requirements. In the spatially-resolved spectrum, shown schematically here, the two spectral peaks are separated by twice their individual bandwidths. A quadratic test phase is assumed for RP, AP and TP.} 
\label{concept}
\end{figure}

The first step in our protocol is to spectrally shear the TP by an amount such that the adjacent well-separated spectral components between two sheared copies spectrally overlap (Fig \ref{concept}). This can be achieved by upconversion with an uncharacterized auxiliary pulse (AP) possessing a continuous bandwidth larger than the required shear. In our implementation we use a spatially chirped AP so that the upconverted signal pulse (SP) is also spatially chirped. We will discuss how this might limit the use of this SPICE implementation in the single-shot case.


In the second step, the signal pulse (SP) obtained from this process is made to interfere either spatially or spectrally with a reference pulse (RP). The RP must have sufficient bandwidth to cover the entire spectrum of the SP, though its phase and amplitude need not be known. The interferogram observed is recorded by means of an imaging spectrometer. The first two steps, and an example of such an interferogram are illustrated in Fig.\ref{concept}. The interference pattern between the SP and the RP reads:
\begin{equation}
\begin{aligned}
I\left(\omega,\Omega,x \right)=&\left|\mathcal{E}_{RP}\left(\omega\right) \cdot a_{RP}\left(x\right)\right|^2 + \\
&\left|\mathcal{E}_{SP}\left(\omega - \Omega\right) \cdot a_{SP}\left(x\right)\right|^2+ \\
& 2 \Re\left( \mathcal{E}_{RP}\left(\omega\right) \cdot \mathcal{E}_{SP}^*\left(\omega - \Omega\right) \cdot \mathcal{A}\left(x\right)\right),
\end{aligned}
\label{inter}
\end{equation}
where $ \mathcal{E}_i \left(\omega\right)$ and $ a_j\left(x\right)$ are the complex valued frequency and spatially dependant components of the electric fields of pulses $\it{i}$ and $\it{j}$ respectively and $\mathcal{A}\left(x\right) = a_{RP}\left(x\right)\cdot a_{SP}^*\left(x\right)$. This term includes any phase offset introduced by the auxiliary field and, differently from other techniques, it can be accounted for in our algorithm. The spatial chirp of the AP is arranged so that the spatial coordinate $x$ and the shear $\Omega$ of the SP have a simple linear relation: $\Omega= \alpha\cdot x$, where the constant $\alpha$ can be obtained via simple calibration of the apparatus.
The phase $\Phi_{AP} (\Omega)$ of $\mathcal{A} (x)$ varies with the shear, since $\Omega $ is itself a function of $x$. This introduces a complication that demands a new approach to extracting the phase of the TP from the measured interferogram.

For any frequency $\omega_0$ in the domain circumscribed by the red rectangle in Fig.1 we can isolate two interferograms, each associated with a different spectral peak, which are spatially sheared by an amount $x_1 (\Omega_1)$ and $x_2 (\Omega_2)$ respectively. In this frequency range it is meaningful to compare the interferometric phases:

\begin{equation}
\begin{aligned}
\Gamma_1\left(\omega_0\right) &= \phi_{SP1}\left(\omega_0\right)-\phi_{RP}\left(\omega_0\right) - \Phi_{AP}\left(\Omega_1\right)&\\ 
\Gamma_2\left(\omega_0\right) &= \phi_{SP2}\left(\omega_0\right)-\phi_{RP}\left(\omega_0\right) - \Phi_{AP}\left(\Omega_2\right),\\
\label{int2}
\end{aligned}
\end{equation}
where $\phi_{SP1}$ and $\phi_{SP2}$ are the phases of the spectral peaks within the SP. If $\Phi_{AP}\left(\Omega_1\right) = \Phi_{AP}\left(\Omega_2\right) = 0$, it would be trivial to obtain the relative phase between the two spectral peaks by taking the difference of the two measurements, and the spectral phase could be retrieved by adopting a standard SPIDER concatenation algorithm. However, this is not the case here and more sophisticated algorithms must be employed. 


The third and final step is the retrieval of the spectral phase of the TP using the recently developed MICE algorithm. The MICE algorithm is an iterative phase reconstruction algorithm that exploits the redundancy present in multi-shearing interferometry to unambiguously separate the fields. It minimizes the error
\begin{equation}
Err = \sum_{j,k}\left|AC_{j,k} - \mathcal{E}_{TP}\left(\omega_j -\Omega_k\right)\cdot\mathcal{E}_{RP}\left(\omega_j\right)\cdot \mathcal{A}\left(\Omega_k\right) \right|^2,
\label{error}
\end{equation}

where $AC_{j,k}$ is the last term of \eqref{inter}, obtained by Fourier filtering using the Takeda algorithm \cite{Takeda82}.
Minimizing \eqref{error} with respect to the three fields leads to the following equations 
\begin{equation}
\begin{aligned}
\mathcal{E}_{RP}\left(\omega_j\right) &= \frac{\sum_k AC^{meas}_{j,j-k}\cdot \mathcal{E}_{TP}\left(\omega_j-\Omega_k\right)\cdot \mathcal{A}^*\left(\Omega_k\right)}{\sum_k \left|\mathcal{E}_{TP}\left(\omega_j-\Omega_k\right)\cdot \mathcal{A}\left(\Omega_k\right)\right|^2} \\&\\
\mathcal{E}_{TP}^*\left(\omega_j\right) &= \frac{\sum_k AC^{meas}_{j+k,j}\cdot \mathcal{E}_{RP}\left(\omega_j+\Omega_k\right)\cdot \mathcal{A}^*\left(\Omega_k\right)}{\Sigma_k \left|\mathcal{E}_{RP}\left(\omega_j+\Omega_k\right)\cdot \mathcal{A}\left(\Omega_k\right)\right|^2}
\\&\\
\mathcal{A}\left(\Omega_k\right) &= \frac{\sum_j AC^{meas}_{j,j-k}\cdot \mathcal{E}_{TP}^*\left(\omega_j-\Omega_k\right)\mathcal{E}_{RP}\left(\omega_j\right)}{\sum_j \left|\mathcal{E}_{TP}\left(\omega_j-\Omega_k\right)\mathcal{E}_{RP}\left(\omega_j\right)\right|^2}
\end{aligned}
\label{MICE}
\end{equation}
which are straightforward to solve using an iterative routine. The redundancy of the data encoding the phase of the TP in the multishear arrangement makes a least-squares fitting algorithm highly efficient and robust to noise in the measurement. and \eqref{MICE} enables the simultaneous estimation of the TP, AP and the RP fields, hence solving the difficulties encountered by the standard SPIDER retrieval approach. Furthermore, MICE does not suffer when using a very broad reference, for there is no need for it to be reconstructed in advance, avoiding propagation of any RP reconstruction error in the final result. A direct measurement of the spectra is needed to resolve the ambiguities in MICE, and can also be useful as an additional constraint. 

As the spectral phase is retrieved by observing a frequency-resolved spatial interferogramme, SPICE is sensitive to the presence of space-time coupling (STC) in either the original beams, or due to the upconversion process. If the STC phase  $\Phi_{STC}= \alpha \left(\omega-\omega_0\right)\left(\Omega-\Omega_0\right)$ is not calibrated, the algorithm mistakenly interprets it as additional quadratic phase on the three fields. When STC is present, the phase of the AC band reads:
\begin{equation}
\begin{aligned}
\Phi_{AC}\left(\omega,\Omega\right)=&\Phi_{RP}\left(\omega\right)+\Phi_{TP}\left(\omega-\Omega\right)\\
&+\Phi_{AP}\left(\Omega\right)+\Phi_{STC}\left(\omega,\Omega\right).
\end{aligned}
\end{equation}
where the last phase term, responsible for STC, is in the form:
\begin{equation}
\Phi_{STC} \left(\omega,\Omega\right) = \gamma\left(\omega-\omega_0\right) \Omega,
\label{STC}
\end{equation}
which produces one term depending exclusively on $\Omega$, which would be assigned as a group delay to the AP only, and one depending on the product $\omega\Omega$. This latter is what effectively prevents the algorithm to successfully retrieve the phases. In fact the term can be decomposed in terms of $\omega, \Omega$, and $\omega-\Omega$ as:
\begin{equation}
\label{equa}
\gamma\omega\Omega =\frac{\gamma}{2}\left( \omega^2 + \Omega^2 -\left(\omega-\Omega\right)^2\right), 
\end{equation}
hence introducing a second order term in each of the three variables. Since the MICE algorithm decomposes the RP, AP and TP fields in terms of $\omega, \Omega$, and $\omega-\Omega$, respectively, it is impossible to assess whether between the second order terms here are the actual GDD of each field or an artefact from the STC. This may constitutes a limitation for this technique, however, since the effect of the STC is described by \eqref{equa}, one can take it into account, if the coefficient $\gamma$ is known, by simply subtracting $\gamma$ to the value of the reconstructed GDD. This implies that a successful implementation of SPICE demands a precalibration on a gapless beam which could be independently characterised, {\it e.g.} by a commercial SPIDER system, although at the price of a reduces accuracy. Alternative schemes could eliminate this issue by alternatives to the spatially-chirped arrangement; if a temporal arrangement is chosen, this could hamper the single-shot use, however the method would retain all its advantages in retrieving the relative phase between separated peaks.

\begin{figure}[!h]
\includegraphics[width=0.4\paperwidth]{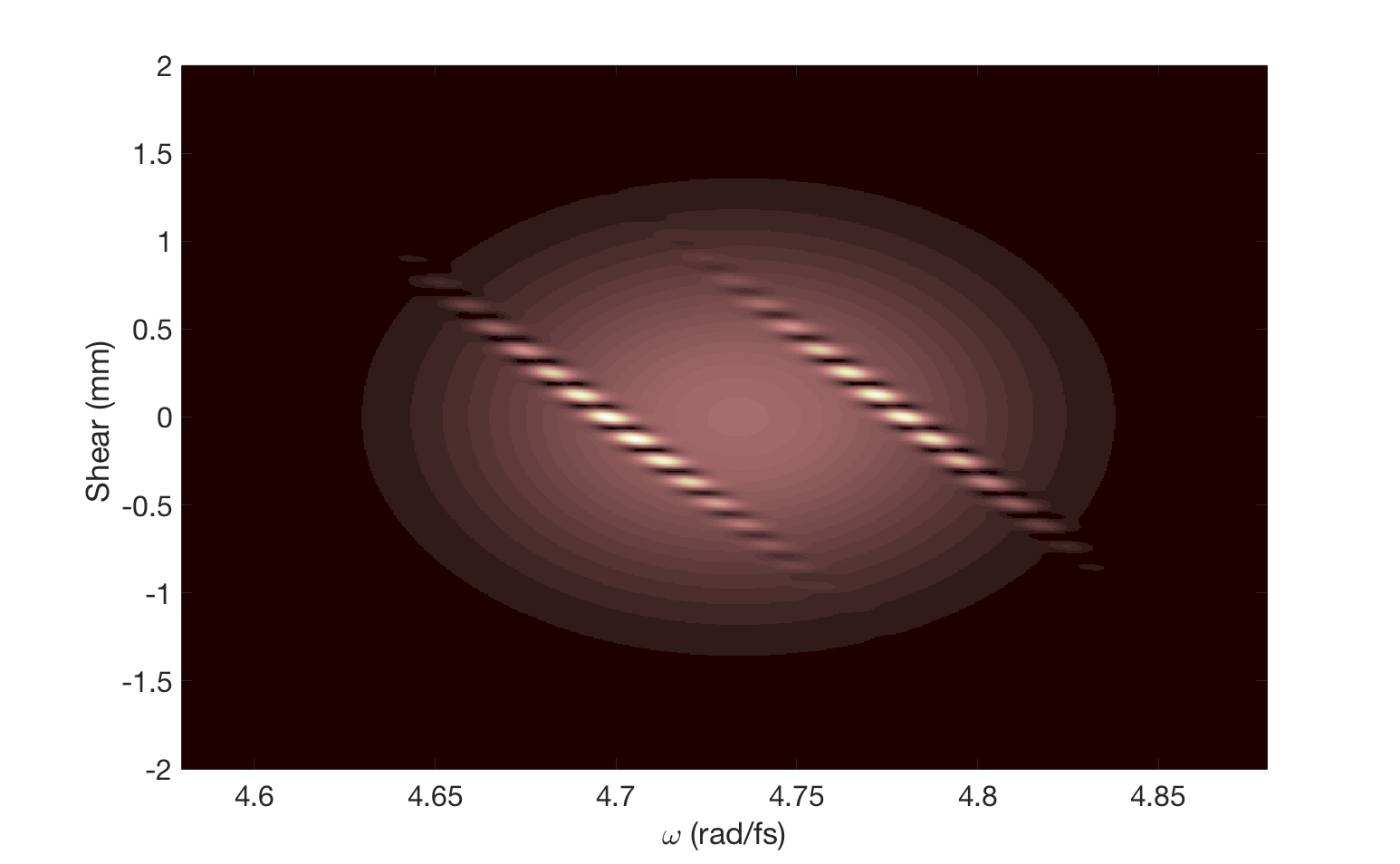}
\caption{ Simulated SPICE interferogram for a filtered TP. The details of the chosen parameters are given in the text.}
\label{interf}
\end{figure}

 \section{Simulations}

We performed simulations with a the filtered SP, in order to check the consistency of the algorithm, and to verify the phase reconstruction. The starting point of the simulation is artificially creating three pulses for the TP, AP, and RP, each with a Gaussian spectrum, and a polynomial spectral phase. The resulting interferogram is of the form given in \eqref {inter}. The first term $\left|\mathcal{E}_{RP}\left(\omega\right) \cdot a_{RP}\left(x\right)\right|^2$ describes the intensity, resolved in frequency and space, of the RP. The second term $\left|\mathcal{E}_{SP}\left(\omega - \Omega\right) \cdot a_{SP}\left(x\right)\right|^2$ provides the same description for the sheared SP. Due to upconversion of the TP with the spatially chirped AP, we obtain a sheared replica of the original input TP spectrum at each spatial position $x$. The last term $2 \Re\left( \mathcal{E}_{RP}\left(\omega\right) \cdot \mathcal{E}_{SP}^*\left(\omega - \Omega\right) \cdot \mathcal{A}\left(x\right)\right)$ is responsible for the spatial fringes, which are then influenced by the three input fields TP, AP, and RP. 

\begin{figure}[h]
\includegraphics[width=0.4\paperwidth]{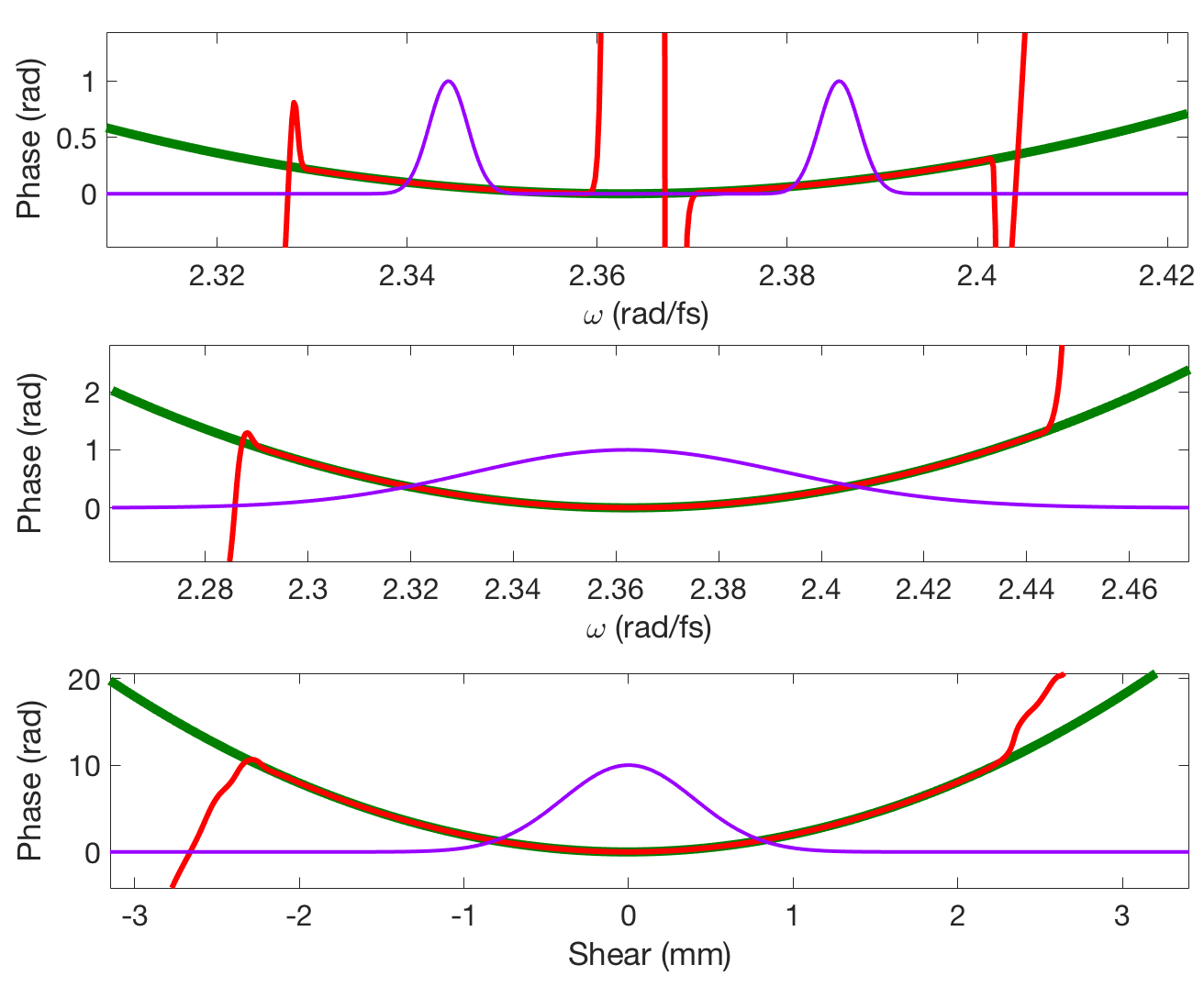}
\caption{Theoretical (green) phase , reconstructed (red ) phase, and spectrum (purple) for the TP (top panel) RP (middle panel) AP (lower panel)}
\label{sim_2}
\end{figure}

In our code, the frequency and shear scale were chosen to be similar to those we employed in our experiment. For the example shown here, the phases of the TP and RP are chosen to be quadratic with GDD= 400 fs$^2$/rad and the phase for the AP is GDD = 4 rad/mm$^2$. The use of different units for the AP is justified in the fact that its spectrum is mapped onto the position scale, therefore, both amplitude and phase will be expressed as a function of the shear $x$. The TP spectrum is filtered so that there are two spectral gaussian components which have $\sigma$ = 10 px and the spectral gap is $\simeq$ 120 px wide. The interferogram so obtained is shown in Fig. \ref{interf}. As in the standard Takeda algorithm~\cite{Takeda82}, a 2-dimensional FFT is performed on the interferogram, the sideband isolated and Fourier-transformed, leading to the reconstruction of the spectral phases via the MICE equations \eqref{MICE}. 

The retrieved spectral phases for the TP, RP and AP  are shown in Fig. \ref{sim_2} together with their expected values. In the first panel of Fig. \ref{sim_2} the relative phase between the two peaks is correctly reconstructed despite the presence of a gap.  In the general case, the spectrum of the RP could happen to span a larger region than the one where interference occurs: in such a case, the reconstructed phase will only be reliable in the bandwidth spanned by the interference, but this would not hamper the reconstruction of the full TP phase. 
 
\begin{figure}[h!]
\includegraphics[width=0.4\paperwidth]{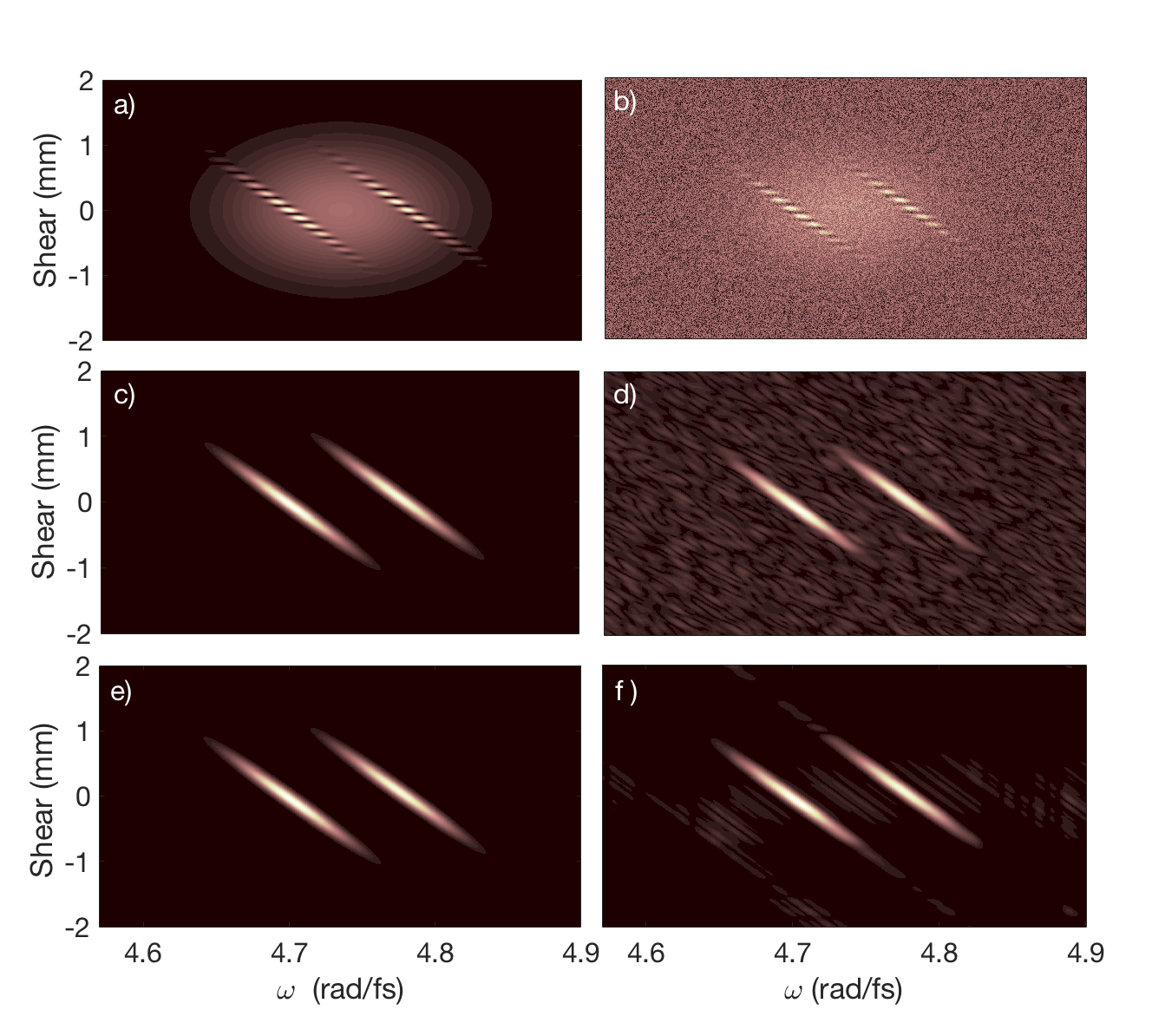}
\caption{  Left column: (a) interferogram , (c) AC side band, (e) retrieved AC side band in the absence of noise; Right column: (b) interferogram, (d) AC side band, (f) retrieved AC side band with 50$\%$ additive noise.}
\label{sim_1}
\end{figure}

The redundancy on the phase information grants to MICE a good robustness against the presence of noise: in Fig. \ref{sim_1} the interferogram, the simulated AC, and the reconstructed AC are shown for 0$\%$ and 50$\%$ additive noise level (left and right column). The sideband, which bears the characteristic two-lobed structure, is well reconstructed even in the presence of high noise.


%

\section{Experiment}

\begin{figure}[!h]
\centering
\includegraphics[width=0.4 \textwidth]{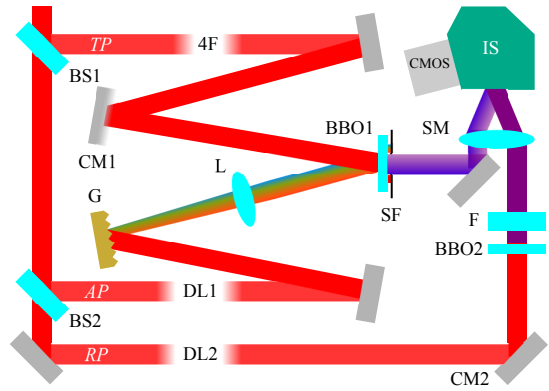}
\caption{Schematic of the experiment: $BS_i$: (PBS+HWP system); 4F: 4f pulse shaper; $DL_i$: delay line; $CM_i$: cylindrical mirror; G: diffraction grating; L: convex lens; BBO: beta-barium borate crystal; SF: spatial filter; F: high-pass filter; SM: spherical mirror; IS: imaging spectrometer; CMOS: CMOS camera.  Distances and angles have been exaggerated for clarity.}
\label{setup}
\end{figure}

We have demonstrated the SPICE experimentally. A beam of 160 mW average power was picked off from a Ti:sapphire chirped-pulsed amplifier delivering 30 fs pulses at a 1 kHz repetition-rate. The pick-off was divided into three beams (TP, AP, and RP) by a series of half wave plates (HWP) and polarizing beam splitters (PBS) to adjust the power as needed. The experimental setup (Fig. \ref{setup}) has been designed to test the method, and includes the facility to create TPs with adjustable spectral gaps. Importantly, the TP, AP and RP need not be derived from the same source, so long as they are synchronized, since their phases are all estimated by the MICE algorithm without assumptions. This means that the power of the TP beam need not be large - since it does not need to contribute to the other beams. This makes SPICE very sensitive to low energy pulses. Of course, the test arrangement used here for convenience does not have this feature, and we used a TP of 40 $\mu J$ energy.

In the TP arm, a folded 4f line (4F) shaped the beam using a removable tunable mask positioned in the Fourier plane. This allowed the generation of pulses with spectral gaps of different widths and could be removed to compare the results with and without a gap present. The mask was manipulated to produce a 12.9 nm gap between two peaks each with a 4 nm spectral support; this scenario would cause common reconstruction methods to fail. The TP was then focused using a cylindrical mirror (CM1) with f = 300 mm into a 100 $\mu m$ thick type-II BBO crystal (BBO1). The AP was spatially-chirped using a diffraction grating (G) and collimated with a spherical lens (L) with focal length f = 60 mm into BBO1 as well. This choice of focusing produces the same spot size for the TP and AP at the focus, guaranteeing complete spatial overlap. 

\begin{figure}[!h]
\includegraphics[width=0.4\paperwidth]{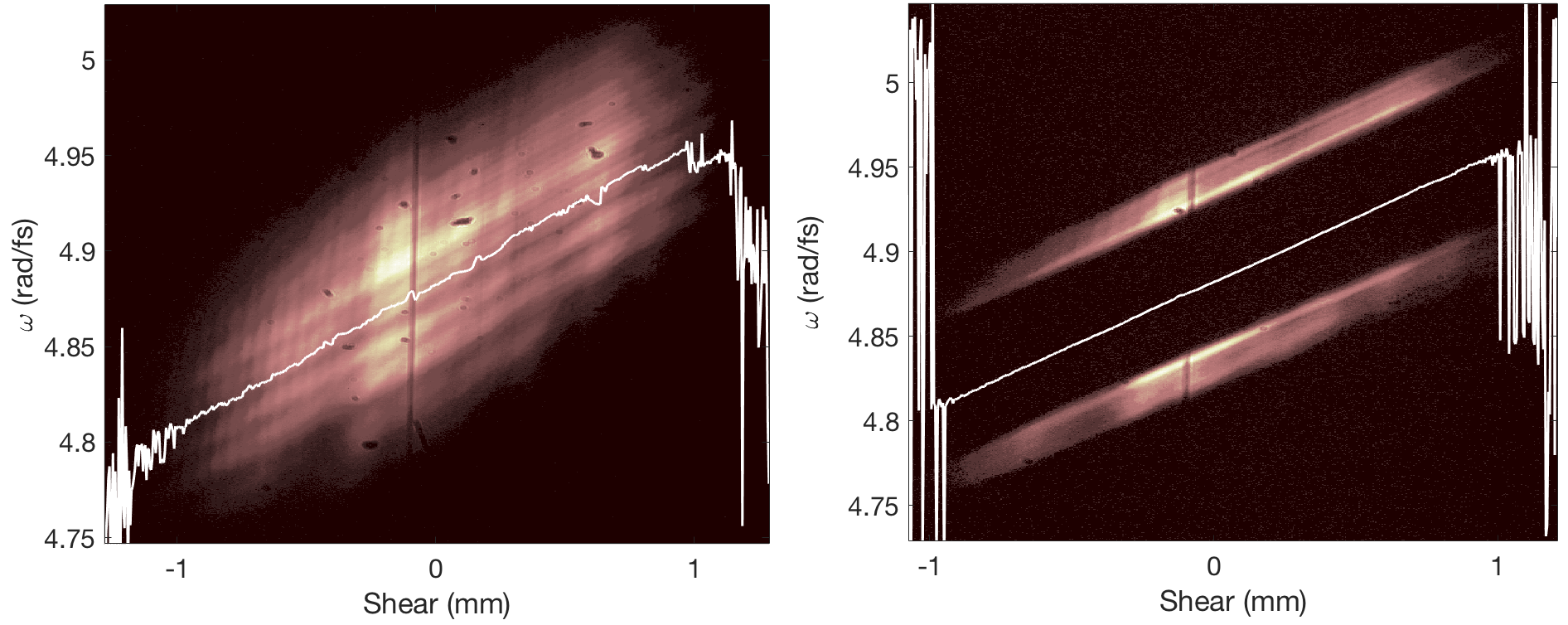}
\caption{Shear calibration of the SP for the unfiltered TP (left panel) and  filtered TP (right panel)}
\label{exp_11}
\end{figure}

The two beams arrive on the crystal in a non-collinear geometry, with angles $\theta_{TP} = 10\degree$ and $\theta_{AP} = 6\degree$ respectively. This allows the phase-matching function to generate sum-frequencies across the whole spectrum of the TP with all frequencies of the AP (at least) spanning the gap. A large phase-matching bandwidth is crucial which must ensure that the spectral gap in the SP is bridged. Furthermore, since shearing is implemented by upconversion, the spatial chirp rate must be sufficiently large that the spectrum of the AP at any given position is sufficiently narrow for spectral features not to be washed out. In our arrangement this corresponds to a chirp rate $\alpha = -50.2 $ mrad/fs/mm; this is mostly imposed by the grating. Since our setup is based on SEA-CAR SPIDER, this value is retrieved directly by recording, together with the interferogram, the SP and RP spatially-resolved spectra separately, as shown in Fig.~\eqref{exp_11}. We notice that a direct measurement of the spectrum is required anyway, since MICE is prone to a exponential error in the retrieved the spectra $e^{(\beta1+i\beta2)\omega}$~\cite{BourassinBouchet2013}.

The upconversion system must be dimensioned to accomodate the needed bandwidth in the transversal dimension of the upconversion crystal. The two beams are synchronized on the crystal by controlling their relative delay by means of a delay line (DL1) on the AP path. The RP is focused with a cylindrical mirror (CM2) of f =300 mm into a second 100 $\mu m$ thick BBO crystal (BBO2). The RP and the SP are then spatially interfered on the entrance slit of an imaging spectrometer (IS) using a $2''$-diameter, f=200 mm spherical mirror (SM). Interferograms are then recorded using a CMOS camera.

\begin{figure}[!h]
\centering
\includegraphics[width=0.45 \textwidth]{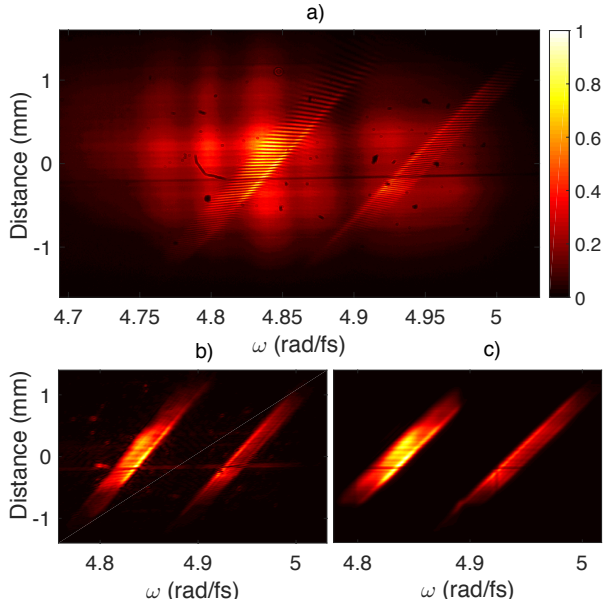}
\caption{Raw data. (a) SPICE interferogram recorded for a TP with a 12.9 nm gap separating two peaks each with a spectral support of 4 nm. (b) The AC components of (a). (c) The AC components of the interferogram calculated from the reconstructed fields. The MICE algorithm converges to this solution after just 10 iterations.}
\label{data_1}
\end{figure}

We measured test pulse fields both unfiltered and filtered with a spectral gap,  with the former serving as a calibration of the STC present in our setup. This can be traced back mostly to two elements: first, the non-collinear geometry for phase-matching the upconversion of the TP with the AP; second, the 4-f line used to shape the spectral gap. The latter is by no means intrinsic to the method, and it is merely an instrumental artefact. The former is mostly dictated by requirements on the upconversion bandwidth. In our experiment, we perform a SPICE measurement on the unfiltered spectrum, and then we perform a separate phase measurement with a commercial SPIDER device. The difference between the two retrieved phases, attributed to STC, is extracted and used as a calibration for the phase of the filtered TP.

The interferogram measured with the imaging spectrometer, averaged over 80ms (i.e. 80 pulses) is shown in Fig. \ref{data_1}(a). The AC components of this and the analogous interferogram calculated from the reconstructed fields are shown in \ref{data_1}(b) and (c). The reconstruction was achieved after just 10 iterations with a MICE error of $0.22\%$, demonstrating the rapid convergence of the MICE algorithm and the high quality of the reconstruction.

\begin{figure}[htp!]
\centering
\includegraphics[width=0.47 \textwidth]{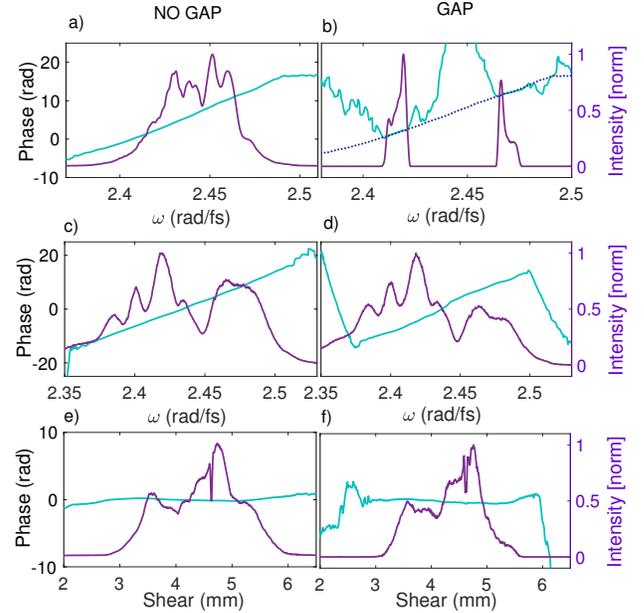}
\caption{Results. The reconstructed spectral phase function (blue) and amplitudes (red) with and without the gap present. (a-b) TP (c-d) RP (e-f) AP. The key result is highlighted in (b), where the reconstructed phase for the filtered TP is overlapped to that in a) (in cyan), showing an excellent agreement. The phase in (c-d) is only reconstructed where there is interference with the SP, which is narrower in (d), resulting in the reconstruction of a smaller portion of the RP phase. }
\label{results}
\end{figure}

The spectral phases and amplitudes obtained for the TP, RP and AP are reported in Fig. \ref{results}, with the same STC phase subtracted. The key result is shown in Fig. \ref{results} (b), where the reconstructed spectral phase function in the regions of the two spectral peaks is in close agreement with that of the unfiltered TP (dashed in blue). 

\begin{figure}[htbp!]
\centering
\includegraphics[width=0.5 \textwidth]{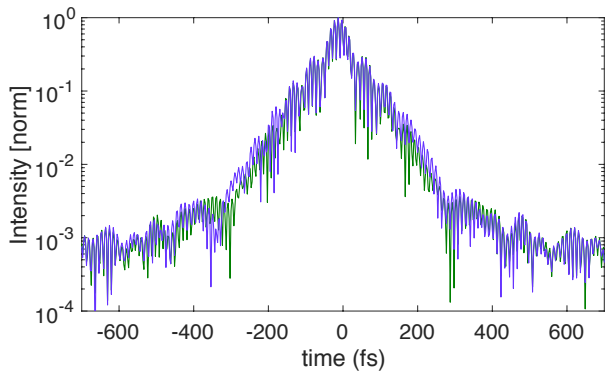}
\caption{ Comparison between the temporal profiles of the pulses $P_1$(violet) and $P_2$(green) in logaritmic scale, obtained using the spectral amplitude of Fig. \ref{results} (b) and the phases shown in Fig. \ref{results} (a) and (b) respectively. This shows that the phase estimated using SPICE are faithful to those of the input unfiltered pulses.}
\label{temporal}
\end{figure}

To test this we plot in Fig. \ref{temporal} the temporal profile estimated from the Fourier Transform of $P_1=S_{gap}\cdot e^{\left(i\Phi_{gap}\right)}$ and $P_2=S_{gap}\cdot e^{\left(i\Phi_{no-gap}\right)}$, where $S_{gap}$ is the spectral amplitude of the filtered TP, and $\Phi_{gap,no-gap}$ are the phases of the filtered and unfiltered TP respectively. These temporal profiles show an overall duration dictated by the full width of the spectrum, modulated by a fast oscillation whose period is linked to the presence of the spectral gap as well as to the chirp. We can observe in Fig. \ref{temporal} a very good agreement between the two temporal profiles wherever the intensity is a substantial fraction of the peak value. This is relfected in a RMSE of $0.57\%$, demonstrating how the relative phase between the two peaks has been correctly reconstructed.

We have demonstrated a new interferometric technique capable of correctly reconstructing the spectral phase of pulses with well-separated spectral components. The new method solves a long-standing challenge for pulse characterisation, and our approach is robust against noise, as well as being scalable to many gaps and large bandwidths. This technique exploits the MICE algorithm to simultaneously reconstruct multiple electric fields.
Since the TP, AP, and RP can be independent, and the spectral phases of none of the pulses need to be known a priori, nonlinear processes such as self-phase modulation or supercontinuum generation can be used to broaden the pulses in order to provide the bandwidth that spans the specral gaps. This makes it possible to employ our method to measure very complex broadband pulses. 
For instance, this can be effective in overcoming the limitations of detector bandwidth:  the UV components could be shifted to the visible via down-conversion, and, simultaneously the IR components could be shifted into the visible via up-conversion. 
We are confident that, as for existing pulse characterization techniques, the sensitivity and versatility of this method can be further refined by incremental improvements to the experimental setup:  alternative arrangements by which the effect of the STC are currently under development.

\section*{Funding}
\label{sec: Funding Information}
This work has been supported by the Engineering and Physical Sciences Research Council (EPSRC) under grants EP/H000178/1 and EP/L015137/1, and by the European Commission, grant no 665148 QCUMBER.

\bibliography{references.bib}

\begin{thebibliography}{20}%
\makeatletter
\providecommand \@ifxundefined [1]{%
 \@ifx{#1\undefined}
}%
\providecommand \@ifnum [1]{%
 \ifnum #1\expandafter \@firstoftwo
 \else \expandafter \@secondoftwo
 \fi
}%
\providecommand \@ifx [1]{%
 \ifx #1\expandafter \@firstoftwo
 \else \expandafter \@secondoftwo
 \fi
}%
\providecommand \natexlab [1]{#1}%
\providecommand \enquote  [1]{``#1''}%
\providecommand \bibnamefont  [1]{#1}%
\providecommand \bibfnamefont [1]{#1}%
\providecommand \citenamefont [1]{#1}%
\providecommand \href@noop [0]{\@secondoftwo}%
\providecommand \href [0]{\begingroup \@sanitize@url \@href}%
\providecommand \@href[1]{\@@startlink{#1}\@@href}%
\providecommand \@@href[1]{\endgroup#1\@@endlink}%
\providecommand \@sanitize@url [0]{\catcode `\\12\catcode `\$12\catcode
  `\&12\catcode `\#12\catcode `\^12\catcode `\_12\catcode `\%12\relax}%
\providecommand \@@startlink[1]{}%
\providecommand \@@endlink[0]{}%
\providecommand \url  [0]{\begingroup\@sanitize@url \@url }%
\providecommand \@url [1]{\endgroup\@href {#1}{\urlprefix }}%
\providecommand \urlprefix  [0]{URL }%
\providecommand \Eprint [0]{\href }%
\providecommand \doibase [0]{http://dx.doi.org/}%
\providecommand \selectlanguage [0]{\@gobble}%
\providecommand \bibinfo  [0]{\@secondoftwo}%
\providecommand \bibfield  [0]{\@secondoftwo}%
\providecommand \translation [1]{[#1]}%
\providecommand \BibitemOpen [0]{}%
\providecommand \bibitemStop [0]{}%
\providecommand \bibitemNoStop [0]{.\EOS\space}%
\providecommand \EOS [0]{\spacefactor3000\relax}%
\providecommand \BibitemShut  [1]{\csname bibitem#1\endcsname}%
\let\auto@bib@innerbib\@empty
\bibitem [{\citenamefont {Walmsley}\ and\ \citenamefont
  {Dorrer}(2009)}]{Walmsley2009}%
  \BibitemOpen
  \bibfield  {author} {\bibinfo {author} {\bibfnamefont {I.~A.}\ \bibnamefont
  {Walmsley}}\ and\ \bibinfo {author} {\bibfnamefont {C.}~\bibnamefont
  {Dorrer}},\ }\href {\doibase 10.1364/AOP.1.000308} {\bibfield  {journal}
  {\bibinfo  {journal} {Advances in Optics and Photonics}\ }\textbf {\bibinfo
  {volume} {1}},\ \bibinfo {pages} {308} (\bibinfo {year} {2009})}\BibitemShut
  {NoStop}%
\bibitem [{\citenamefont {Krausz}\ and\ \citenamefont
  {Ivanov}(2009)}]{Krausz2009}%
  \BibitemOpen
  \bibfield  {author} {\bibinfo {author} {\bibfnamefont {F.}~\bibnamefont
  {Krausz}}\ and\ \bibinfo {author} {\bibfnamefont {M.}~\bibnamefont
  {Ivanov}},\ }\href {\doibase 10.1103/RevModPhys.81.163} {\bibfield  {journal}
  {\bibinfo  {journal} {Reviews of Modern Physics}\ }\textbf {\bibinfo {volume}
  {81}},\ \bibinfo {pages} {163} (\bibinfo {year} {2009})}\BibitemShut
  {NoStop}%
\bibitem [{\citenamefont {{Potter E. D., Herek L.J., Pedersen S., Liu
  Q.}}(1992)}]{Potter}%
  \BibitemOpen
  \bibfield  {author} {\bibinfo {author} {\bibfnamefont {Z.~A.}\ \bibnamefont
  {{Potter E. D., Herek L.J., Pedersen S., Liu Q.}}},\ }\href@noop {}
  {\bibfield  {journal} {\bibinfo  {journal} {Nature}\ }\textbf {\bibinfo
  {volume} {355}},\ \bibinfo {pages} {66} (\bibinfo {year} {1992})}\BibitemShut
  {NoStop}%
\bibitem [{\citenamefont {Brumer}\ and\ \citenamefont
  {Shapiro}(1986)}]{Brumer1986}%
  \BibitemOpen
  \bibfield  {author} {\bibinfo {author} {\bibfnamefont {P.}~\bibnamefont
  {Brumer}}\ and\ \bibinfo {author} {\bibfnamefont {M.}~\bibnamefont
  {Shapiro}},\ }\href {\doibase 10.1016/S0009-2614(86)80171-3} {\bibfield
  {journal} {\bibinfo  {journal} {Chemical Physics Letters}\ }\textbf {\bibinfo
  {volume} {126}},\ \bibinfo {pages} {541} (\bibinfo {year}
  {1986})}\BibitemShut {NoStop}%
\bibitem [{\citenamefont {Manzoni}\ \emph {et~al.}(2015)\citenamefont
  {Manzoni}, \citenamefont {Mucke}, \citenamefont {Cirmi}, \citenamefont
  {Fang}, \citenamefont {Moses}, \citenamefont {Huang}, \citenamefont {Hong},
  \citenamefont {Cerullo},\ and\ \citenamefont {Kartner}}]{Manzoni2015}%
  \BibitemOpen
  \bibfield  {author} {\bibinfo {author} {\bibfnamefont {C.}~\bibnamefont
  {Manzoni}}, \bibinfo {author} {\bibfnamefont {O.~D.}\ \bibnamefont {Mucke}},
  \bibinfo {author} {\bibfnamefont {G.}~\bibnamefont {Cirmi}}, \bibinfo
  {author} {\bibfnamefont {S.}~\bibnamefont {Fang}}, \bibinfo {author}
  {\bibfnamefont {J.}~\bibnamefont {Moses}}, \bibinfo {author} {\bibfnamefont
  {S.~W.}\ \bibnamefont {Huang}}, \bibinfo {author} {\bibfnamefont {K.~H.}\
  \bibnamefont {Hong}}, \bibinfo {author} {\bibfnamefont {G.}~\bibnamefont
  {Cerullo}}, \ and\ \bibinfo {author} {\bibfnamefont {F.~X.}\ \bibnamefont
  {Kartner}},\ }\href {\doibase 10.1002/lpor.201400181} {\bibfield  {journal}
  {\bibinfo  {journal} {Laser and Photonics Reviews}\ }\textbf {\bibinfo
  {volume} {9}},\ \bibinfo {pages} {129} (\bibinfo {year} {2015})}\BibitemShut
  {NoStop}%
\bibitem [{\citenamefont {Wirth}\ \emph {et~al.}(2011)\citenamefont {Wirth},
  \citenamefont {Hassan}, \citenamefont {Grguras}, \citenamefont {Gagnon},
  \citenamefont {Moulet}, \citenamefont {Luu}, \citenamefont {Pabst},
  \citenamefont {Santra}, \citenamefont {Alahmed}, \citenamefont {Azzeer},
  \citenamefont {Yakovlev}, \citenamefont {Pervak}, \citenamefont {Krausz},\
  and\ \citenamefont {Goulielmakis}}]{Wirth2011}%
  \BibitemOpen
  \bibfield  {author} {\bibinfo {author} {\bibfnamefont {A.}~\bibnamefont
  {Wirth}}, \bibinfo {author} {\bibfnamefont {M.~T.}\ \bibnamefont {Hassan}},
  \bibinfo {author} {\bibfnamefont {I.}~\bibnamefont {Grguras}}, \bibinfo
  {author} {\bibfnamefont {J.}~\bibnamefont {Gagnon}}, \bibinfo {author}
  {\bibfnamefont {A.}~\bibnamefont {Moulet}}, \bibinfo {author} {\bibfnamefont
  {T.~T.}\ \bibnamefont {Luu}}, \bibinfo {author} {\bibfnamefont
  {S.}~\bibnamefont {Pabst}}, \bibinfo {author} {\bibfnamefont
  {R.}~\bibnamefont {Santra}}, \bibinfo {author} {\bibfnamefont {Z.~a.}\
  \bibnamefont {Alahmed}}, \bibinfo {author} {\bibfnamefont {a.~M.}\
  \bibnamefont {Azzeer}}, \bibinfo {author} {\bibfnamefont {V.~S.}\
  \bibnamefont {Yakovlev}}, \bibinfo {author} {\bibfnamefont {V.}~\bibnamefont
  {Pervak}}, \bibinfo {author} {\bibfnamefont {F.}~\bibnamefont {Krausz}}, \
  and\ \bibinfo {author} {\bibfnamefont {E.}~\bibnamefont {Goulielmakis}},\
  }\href {\doibase 10.1126/science.1210268} {\bibfield  {journal} {\bibinfo
  {journal} {Science}\ }\textbf {\bibinfo {volume} {334}},\ \bibinfo {pages}
  {195} (\bibinfo {year} {2011})}\BibitemShut {NoStop}%
\bibitem [{\citenamefont {McCracken}\ \emph {et~al.}(2012)\citenamefont
  {McCracken}, \citenamefont {Sun}, \citenamefont {Leburn},\ and\ \citenamefont
  {Reid}}]{McCracken2012}%
  \BibitemOpen
  \bibfield  {author} {\bibinfo {author} {\bibfnamefont {R.~A.}\ \bibnamefont
  {McCracken}}, \bibinfo {author} {\bibfnamefont {J.}~\bibnamefont {Sun}},
  \bibinfo {author} {\bibfnamefont {C.~G.}\ \bibnamefont {Leburn}}, \ and\
  \bibinfo {author} {\bibfnamefont {D.~T.}\ \bibnamefont {Reid}},\ }\href
  {\doibase 10.1364/OE.20.016269} {\bibfield  {journal} {\bibinfo  {journal}
  {Optics Express}\ }\textbf {\bibinfo {volume} {20}},\ \bibinfo {pages}
  {16269} (\bibinfo {year} {2012})}\BibitemShut {NoStop}%
\bibitem [{\citenamefont {Huang}\ \emph {et~al.}(2011)\citenamefont {Huang},
  \citenamefont {Cirmi}, \citenamefont {Moses}, \citenamefont {Hong},
  \citenamefont {Bhardwaj}, \citenamefont {Birge}, \citenamefont {Chen},
  \citenamefont {Li}, \citenamefont {Eggleton}, \citenamefont {Cerullo},\ and\
  \citenamefont {K{\"{a}}rtner}}]{Huang2011}%
  \BibitemOpen
  \bibfield  {author} {\bibinfo {author} {\bibfnamefont {S.-W.}\ \bibnamefont
  {Huang}}, \bibinfo {author} {\bibfnamefont {G.}~\bibnamefont {Cirmi}},
  \bibinfo {author} {\bibfnamefont {J.}~\bibnamefont {Moses}}, \bibinfo
  {author} {\bibfnamefont {K.-H.}\ \bibnamefont {Hong}}, \bibinfo {author}
  {\bibfnamefont {S.}~\bibnamefont {Bhardwaj}}, \bibinfo {author}
  {\bibfnamefont {J.~R.}\ \bibnamefont {Birge}}, \bibinfo {author}
  {\bibfnamefont {L.-J.}\ \bibnamefont {Chen}}, \bibinfo {author}
  {\bibfnamefont {E.}~\bibnamefont {Li}}, \bibinfo {author} {\bibfnamefont
  {B.~J.}\ \bibnamefont {Eggleton}}, \bibinfo {author} {\bibfnamefont
  {G.}~\bibnamefont {Cerullo}}, \ and\ \bibinfo {author} {\bibfnamefont
  {F.~X.}\ \bibnamefont {K{\"{a}}rtner}},\ }\href {\doibase
  10.1038/nphoton.2011.140} {\bibfield  {journal} {\bibinfo  {journal} {Nature
  Photonics}\ }\textbf {\bibinfo {volume} {5}},\ \bibinfo {pages} {475}
  (\bibinfo {year} {2011})}\BibitemShut {NoStop}%
\bibitem [{\citenamefont {Austin}\ \emph {et~al.}(2010)\citenamefont {Austin},
  \citenamefont {Witting},\ and\ \citenamefont {Walmsley}}]{Austin2010b}%
  \BibitemOpen
  \bibfield  {author} {\bibinfo {author} {\bibfnamefont {D.~R.}\ \bibnamefont
  {Austin}}, \bibinfo {author} {\bibfnamefont {T.}~\bibnamefont {Witting}}, \
  and\ \bibinfo {author} {\bibfnamefont {I.~A.}\ \bibnamefont {Walmsley}},\
  }\href {\doibase 10.1364/OL.35.001971} {\bibfield  {journal} {\bibinfo
  {journal} {Optics letters}\ }\textbf {\bibinfo {volume} {35}},\ \bibinfo
  {pages} {1971} (\bibinfo {year} {2010})}\BibitemShut {NoStop}%
\bibitem [{\citenamefont {Hirasawa}\ \emph {et~al.}(2002)\citenamefont
  {Hirasawa}, \citenamefont {Nakagawa}, \citenamefont {Yamamoto}, \citenamefont
  {Morita}, \citenamefont {Shigekawa},\ and\ \citenamefont
  {Yamashita}}]{Hirasawa2002}%
  \BibitemOpen
  \bibfield  {author} {\bibinfo {author} {\bibfnamefont {M.}~\bibnamefont
  {Hirasawa}}, \bibinfo {author} {\bibfnamefont {N.}~\bibnamefont {Nakagawa}},
  \bibinfo {author} {\bibfnamefont {K.}~\bibnamefont {Yamamoto}}, \bibinfo
  {author} {\bibfnamefont {R.}~\bibnamefont {Morita}}, \bibinfo {author}
  {\bibfnamefont {H.}~\bibnamefont {Shigekawa}}, \ and\ \bibinfo {author}
  {\bibfnamefont {M.}~\bibnamefont {Yamashita}},\ }\href {\doibase
  10.1007/s00340-002-0891-y} {\bibfield  {journal} {\bibinfo  {journal}
  {Applied Physics B: Lasers and Optics}\ }\textbf {\bibinfo {volume} {74}},\
  \bibinfo {pages} {225} (\bibinfo {year} {2002})}\BibitemShut {NoStop}%
\bibitem [{\citenamefont {Morita}\ \emph {et~al.}(2002)\citenamefont {Morita},
  \citenamefont {Hirasawa}, \citenamefont {Karasawa}, \citenamefont {Kusaka},
  \citenamefont {Nakagawa}, \citenamefont {Yamane}, \citenamefont {Li},
  \citenamefont {Suguro},\ and\ \citenamefont {Yamashita}}]{Morita2002}%
  \BibitemOpen
  \bibfield  {author} {\bibinfo {author} {\bibfnamefont {R.}~\bibnamefont
  {Morita}}, \bibinfo {author} {\bibfnamefont {M.}~\bibnamefont {Hirasawa}},
  \bibinfo {author} {\bibfnamefont {N.}~\bibnamefont {Karasawa}}, \bibinfo
  {author} {\bibfnamefont {S.}~\bibnamefont {Kusaka}}, \bibinfo {author}
  {\bibfnamefont {N.}~\bibnamefont {Nakagawa}}, \bibinfo {author}
  {\bibfnamefont {K.}~\bibnamefont {Yamane}}, \bibinfo {author} {\bibfnamefont
  {L.}~\bibnamefont {Li}}, \bibinfo {author} {\bibfnamefont {A.}~\bibnamefont
  {Suguro}}, \ and\ \bibinfo {author} {\bibfnamefont {M.}~\bibnamefont
  {Yamashita}},\ }\href {\doibase 10.1088/0957-0233/13/11/307} {\bibfield
  {journal} {\bibinfo  {journal} {Measurement Science and Technology}\ }\textbf
  {\bibinfo {volume} {13}},\ \bibinfo {pages} {1710} (\bibinfo {year}
  {2002})}\BibitemShut {NoStop}%
\bibitem [{\citenamefont {Linden}\ \emph {et~al.}(1998)\citenamefont {Linden},
  \citenamefont {Giessen},\ and\ \citenamefont {Kuhl}}]{Linden1998}%
  \BibitemOpen
  \bibfield  {author} {\bibinfo {author} {\bibfnamefont {S.}~\bibnamefont
  {Linden}}, \bibinfo {author} {\bibfnamefont {H.}~\bibnamefont {Giessen}}, \
  and\ \bibinfo {author} {\bibfnamefont {J.}~\bibnamefont {Kuhl}},\ }\href
  {\doibase 10.1002/(SICI)1521-3951(199803)206:1<119::AID-PSSB119>3.0.CO;2-X}
  {\bibfield  {journal} {\bibinfo  {journal} {Physica Status Solidi (B)}\
  }\textbf {\bibinfo {volume} {206}},\ \bibinfo {pages} {119} (\bibinfo {year}
  {1998})}\BibitemShut {NoStop}%
\bibitem [{\citenamefont {Keusters}\ \emph {et~al.}(2003)\citenamefont
  {Keusters}, \citenamefont {Tan}, \citenamefont {O'Shea}, \citenamefont
  {Zeek}, \citenamefont {Trebino},\ and\ \citenamefont
  {Warren}}]{Keusters2003}%
  \BibitemOpen
  \bibfield  {author} {\bibinfo {author} {\bibfnamefont {D.}~\bibnamefont
  {Keusters}}, \bibinfo {author} {\bibfnamefont {H.-S.}\ \bibnamefont {Tan}},
  \bibinfo {author} {\bibfnamefont {P.}~\bibnamefont {O'Shea}}, \bibinfo
  {author} {\bibfnamefont {E.}~\bibnamefont {Zeek}}, \bibinfo {author}
  {\bibfnamefont {R.}~\bibnamefont {Trebino}}, \ and\ \bibinfo {author}
  {\bibfnamefont {W.~S.}\ \bibnamefont {Warren}},\ }\href {\doibase
  10.1364/JOSAB.20.002226} {\bibfield  {journal} {\bibinfo  {journal} {Journal
  of the Optical Society of America B}\ }\textbf {\bibinfo {volume} {20}},\
  \bibinfo {pages} {2226} (\bibinfo {year} {2003})}\BibitemShut {NoStop}%
\bibitem [{\citenamefont {Seifert}\ \emph {et~al.}(2004)\citenamefont
  {Seifert}, \citenamefont {Stolz},\ and\ \citenamefont
  {Tasche}}]{Seifert2004}%
  \BibitemOpen
  \bibfield  {author} {\bibinfo {author} {\bibfnamefont {B.}~\bibnamefont
  {Seifert}}, \bibinfo {author} {\bibfnamefont {H.}~\bibnamefont {Stolz}}, \
  and\ \bibinfo {author} {\bibfnamefont {M.}~\bibnamefont {Tasche}},\ }\href
  {\doibase 10.1364/JOSAB.21.001089} {\bibfield  {journal} {\bibinfo  {journal}
  {Journal of the Optical Society of America B}\ }\textbf {\bibinfo {volume}
  {21}},\ \bibinfo {pages} {1089} (\bibinfo {year} {2004})}\BibitemShut
  {NoStop}%
\bibitem [{\citenamefont {Seifert}\ and\ \citenamefont
  {Stolz}(2009)}]{Seifert2009}%
  \BibitemOpen
  \bibfield  {author} {\bibinfo {author} {\bibfnamefont {B.}~\bibnamefont
  {Seifert}}\ and\ \bibinfo {author} {\bibfnamefont {H.}~\bibnamefont
  {Stolz}},\ }\href {\doibase 10.1088/0957-0233/20/1/015303} {\bibfield
  {journal} {\bibinfo  {journal} {Measurement Science and Technology}\ }\textbf
  {\bibinfo {volume} {20}},\ \bibinfo {pages} {015303} (\bibinfo {year}
  {2009})}\BibitemShut {NoStop}%
\bibitem [{\citenamefont {Wyatt}\ \emph {et~al.}(2013)\citenamefont {Wyatt},
  \citenamefont {Witting}, \citenamefont {Schiavi}, \citenamefont {Fabris},
  \citenamefont {Marangos}, \citenamefont {Tisch},\ and\ \citenamefont
  {Walmsley}}]{ARIES}%
  \BibitemOpen
  \bibfield  {author} {\bibinfo {author} {\bibfnamefont {A.~S.}\ \bibnamefont
  {Wyatt}}, \bibinfo {author} {\bibfnamefont {T.}~\bibnamefont {Witting}},
  \bibinfo {author} {\bibfnamefont {A.}~\bibnamefont {Schiavi}}, \bibinfo
  {author} {\bibfnamefont {D.}~\bibnamefont {Fabris}}, \bibinfo {author}
  {\bibfnamefont {J.~P.}\ \bibnamefont {Marangos}}, \bibinfo {author}
  {\bibfnamefont {J.~W.~G.}\ \bibnamefont {Tisch}}, \ and\ \bibinfo {author}
  {\bibfnamefont {I.~A.}\ \bibnamefont {Walmsley}},\ }\href {\doibase
  10.1109/CLEOE-IQEC.2013.6801150} {\bibfield  {journal} {\bibinfo  {journal}
  {2013 Conference on Lasers and Electro-Optics Europe and International
  Quantum Electronics Conference, CLEO/Europe-IQEC 2013}\ }\textbf {\bibinfo
  {volume} {3}} (\bibinfo {year} {2013}),\
  10.1109/CLEOE-IQEC.2013.6801150}\BibitemShut {NoStop}%
\bibitem [{\citenamefont {Kim}\ \emph {et~al.}(2013)\citenamefont {Kim},
  \citenamefont {Zhang}, \citenamefont {Shiner}, \citenamefont {Schmidt},
  \citenamefont {L{\'{e}}gar{\'{e}}}, \citenamefont {Villeneuve},\ and\
  \citenamefont {Corkum}}]{Kim2013}%
  \BibitemOpen
  \bibfield  {author} {\bibinfo {author} {\bibfnamefont {K.~T.}\ \bibnamefont
  {Kim}}, \bibinfo {author} {\bibfnamefont {C.}~\bibnamefont {Zhang}}, \bibinfo
  {author} {\bibfnamefont {A.~D.}\ \bibnamefont {Shiner}}, \bibinfo {author}
  {\bibfnamefont {B.~E.}\ \bibnamefont {Schmidt}}, \bibinfo {author}
  {\bibfnamefont {F.}~\bibnamefont {L{\'{e}}gar{\'{e}}}}, \bibinfo {author}
  {\bibfnamefont {D.~M.}\ \bibnamefont {Villeneuve}}, \ and\ \bibinfo {author}
  {\bibfnamefont {P.~B.}\ \bibnamefont {Corkum}},\ }\href {\doibase
  10.1038/nphoton.2013.286} {\bibfield  {journal} {\bibinfo  {journal} {Nature
  Photonics}\ }\textbf {\bibinfo {volume} {7}},\ \bibinfo {pages} {1} (\bibinfo
  {year} {2013})}\BibitemShut {NoStop}%
\bibitem [{\citenamefont {Witting}\ \emph {et~al.}(2009)\citenamefont
  {Witting}, \citenamefont {Austin},\ and\ \citenamefont
  {Walmsley}}]{Witting2009a}%
  \BibitemOpen
  \bibfield  {author} {\bibinfo {author} {\bibfnamefont {T.}~\bibnamefont
  {Witting}}, \bibinfo {author} {\bibfnamefont {D.~R.}\ \bibnamefont {Austin}},
  \ and\ \bibinfo {author} {\bibfnamefont {I.~A.}\ \bibnamefont {Walmsley}},\
  }\href {\doibase 10.1364/OL.34.000881} {\bibfield  {journal} {\bibinfo
  {journal} {Optics letters}\ }\textbf {\bibinfo {volume} {34}},\ \bibinfo
  {pages} {881} (\bibinfo {year} {2009})}\BibitemShut {NoStop}%
\bibitem [{\citenamefont {Bourassin-Bouchet}\ \emph {et~al.}(2013)\citenamefont
  {Bourassin-Bouchet}, \citenamefont {Mang}, \citenamefont {Gianani},\ and\
  \citenamefont {Walmsley}}]{BourassinBouchet2013}%
  \BibitemOpen
  \bibfield  {author} {\bibinfo {author} {\bibfnamefont {C.}~\bibnamefont
  {Bourassin-Bouchet}}, \bibinfo {author} {\bibfnamefont {M.~M.}\ \bibnamefont
  {Mang}}, \bibinfo {author} {\bibfnamefont {I.}~\bibnamefont {Gianani}}, \
  and\ \bibinfo {author} {\bibfnamefont {I.~a.}\ \bibnamefont {Walmsley}},\
  }\href {\doibase 10.1364/OL.38.005299} {\bibfield  {journal} {\bibinfo
  {journal} {Optics letters}\ }\textbf {\bibinfo {volume} {38}},\ \bibinfo
  {pages} {5299} (\bibinfo {year} {2013})}\BibitemShut {NoStop}%
\bibitem [{\citenamefont {Takeda}\ \emph {et~al.}(1982)\citenamefont {Takeda},
  \citenamefont {Ina},\ and\ \citenamefont {Kobayashi}}]{Takeda82}%
  \BibitemOpen
  \bibfield  {author} {\bibinfo {author} {\bibfnamefont {M.}~\bibnamefont
  {Takeda}}, \bibinfo {author} {\bibfnamefont {H.}~\bibnamefont {Ina}}, \ and\
  \bibinfo {author} {\bibfnamefont {S.}~\bibnamefont {Kobayashi}},\ }\href
  {\doibase 10.1364/JOSA.72.000156} {\bibfield  {journal} {\bibinfo  {journal}
  {J. Opt. Soc. Am.}\ }\textbf {\bibinfo {volume} {72}},\ \bibinfo {pages}
  {156} (\bibinfo {year} {1982})}\BibitemShut {NoStop}%
\end{thebibliography}%

\end{document}